\documentclass[aps,showpacs,twocolumn]{revtex4}

\usepackage{amsmath}
\usepackage{amssymb}                 
\usepackage{amscd}                   
\usepackage{wasysym}
\usepackage[ansinew]{inputenc}
\usepackage[T1]{fontenc}
\usepackage{ae,aecompl}
\usepackage[dvips]{graphicx}
\DeclareGraphicsExtensions{.eps}

\newcommand{\ri}{{ \rm i }}
\newcommand{\re}{{ \rm e }}
\newcommand{\rd}{{ \rm d }}

\newcommand{\be}{\begin{equation}}
\newcommand{\ee}{\end{equation}}
\newcommand{\nn}{\nonumber}

\begin{document}
\bibliographystyle{apsrev}
\title{Towards a Landau-Zener formula for an interacting Bose-Einstein condensate}
\author{D. Witthaut, E. M. Graefe, and H. J. Korsch}
\email{witthaut@physik.uni-kl.de}
\affiliation{FB Physik, Technische Universit{\"a}t Kaiserslautern,
D-67653 Kaiserslautern, Germany}
\date{\today }

\begin{abstract}
We consider the Landau-Zener problem for a Bose-Einstein condensate
in a linearly varying two-level system, for the full many-particle
system as well and in the mean-field approximation.
The many-particle problem can be solved approximately within an
independent crossings approximation, which yields an explicit
Landau-Zener formula.
\end{abstract}

\pacs{03.75.Lm, 03.65.-w, 73.40.Gk}
\maketitle


\section{Introduction}

During the last years, a lot of work has been devoted to the nonlinear
Landau-Zener problem, which describes a Bose-Einstein condensate (BEC)
in a time-dependent two-state system in the mean-field approximation
\cite{Wu00,Wu03,Liu03}.
As in the celebrated original Landau-Zener scenario, the energy difference
between the two levels is assumed to vary linearly in time.
This situation arises, e.g., for a BEC in a double-well trap or for a BEC
in an accelerated lattice around the edge of the Brillouin zone.
A major question in such a situation is the following: Initially the
two states are energetically well separated and the total population
is in the lower state. Then the energy difference varies linearly in
time, such that the two levels (anti-) cross. Finally the states are
energetically well separated again, however they are just exchanged.
What is the probability of a diabatic time-evolution, i.e. how much of
the initial population remains in the first (diabatic) state ?

In the mean-field approximation, the time evolution is given by the
Gross-Pitaevskii equation
\be
  \ri \frac{\rd}{\rd t}
  \left(\begin{array}{c} \psi_{1} \\ \psi_2 \end{array} \right)
  = \hat H(|\psi_1|^2,|\psi_2|^2,t)
  \left(\begin{array}{c} \psi_{1} \\ \psi_2 \end{array} \right),
\ee
with the nonlinear Hamiltonian
\be
  \hat H(|\psi_1|^2,|\psi_2|^2,t) = \left(\begin{array}{c c}
  \epsilon + g |\psi_1|^2 & v \\ v & -\epsilon + g |\psi_2|^2
  \end{array} \right).
  \label{eqn-ham-nonlin}
\ee
and $\epsilon = \alpha t$. The state vector is normalized to unity,
thus the effective nonlinearity is $g = \bar g N$, where $N$ is number
of particles in the condensate and $\bar g$ is the bare two particle
interaction constant. Throughout this paper we use scaled units
such that $\hbar =  1$.

\begin{figure}[t]
\centering
\includegraphics[width=8cm,  angle=0]{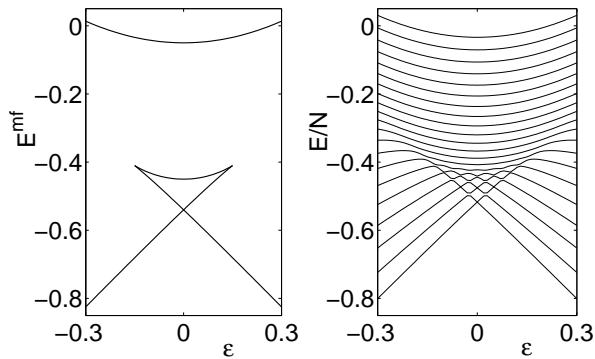}
\caption{\label{fig-levels_mp_mf}
Total energy (\ref{eqn-nonlin-etot}) in the mean-field theory (left)
and eigenenergies the many-particle Hamiltonian (\ref{eqn-mp-hamiltonian}),(right)
for $v = 0.2$, $g = -1$ and $N=20$ particles.}
\end{figure}

The Landau-Zener transition probability is defined as
\be
  P_{\rm LZ}^{\rm mf} = \frac{|\psi_1(t \rightarrow + \infty)|^2}{
  |\psi_1(t \rightarrow - \infty)|^2} \, .
  \label{eqn-plz-mf-def}
\ee
The original linear problem can be solved analytically with different
approaches \cite{Land32,Zene32,Majo32,Stue32}.
This yields the celebrated Landau-Zener formula
\be
 P_{\rm LZ}^{\rm lin}  = \re^{-\pi v^2/\alpha}  \quad \mbox{for} \, g = 0.
 \label{eqn-plz-lin}
\ee
for the probability of a diabatic time evolution.
In the nonlinear case $g < 0$, things get quite complicated
and the Landau-Zener probability is seriously altered.
New nonlinear eigenstates emerge if the nonlinearity exceeds
a critical value $|g| > g_c = 2v$.
A loop develops at the top of the lowest level $\mu(\epsilon)$,
while the total energy
\begin{eqnarray}
  E^{\rm mf} &=& \epsilon (|\psi_1|^2 - |\psi_2|^2) + \frac{g}{2} (|\psi_1|^4 + |\psi_2|^4) \nonumber \\
  && \quad + v (\psi_1^* \psi_2 + \psi_2^* \psi_1)
  \label{eqn-nonlin-etot}
\end{eqnarray}
shows a swallow's tail structure
(cf. the left-hand side of Fig.~\ref{fig-levels_mp_mf}).
The system can evolve adiabatically along this level only up to
the end of the loop - adiabaticity breaks down.
Consequently, the Landau-Zener probability does not vanish even in
the adiabatic limit $\alpha \rightarrow 0$ \cite{Wu00,Wu03}.
For repulsive nonlinearities, $g > 0$ , the situation is just the
other way round: The loop appears in the upper level, thus no adiabatic
evolution is possible in the upper level. In this paper we consider
only the lower level and thus the attractive case $g \le 0$.
These considerations have let to a reformulation of the adiabatic
theorem for nonlinear systems, based on the adiabatic theorem of
classical mechanics \cite{Liu03}.
Note also that the emergence of looped levels was previously
studied for the quantum dimer \cite{Esse95}.

Several approaches were made to derive a nonlinear Landau-Zener
formula for this problem using methods from classical Hamiltonian
mechanics \cite{Zoba00,Liu02}.
For subcritical values of the nonlinearity $|g| < g_c$,
standard methods of classical nonadiabatic corrections yield
good results for the near-adiabatic case ($\alpha / v^2 \ll 1$).
For the case of a rapid passage ($ v^2/\alpha \ll 1$)
one finds a quantitative good approximation using classical
perturbation theory with $v$ being the small parameter
for the subcritical regime as well as for strong nonlinearities,
as long as $g<0$.
Furthermore for strong nonlinearities there is a simple formula which
provides a good approximation for the tunneling probability for an
intermediate range of the parameter $\alpha$. This approximation
fails in the rapid limit as well as in the near adiabatic one.
However, there is no valid approximation in the critical regime
$|g| > g_c = 2v$ for $\alpha \rightarrow 0$ so far. Since
one is interested in the quasiadiabatic dynamic in most
applications this is an important\begin{scriptsize}\end{scriptsize} deficit. Here we present a different
approach which yields good results especially in this region.

Going back to the roots of the problem, we consider the original
many-particle problem of an interacting two-mode boson field
instead of the mean-field theory.
We consider the many-particle Hamiltonian of Bose-Hubbard type,
\begin{eqnarray}
  \hat H(t) &=& \epsilon(t) (\hat n_1 - \hat n_2)
   + v(\hat a_1^\dagger \hat a_2 + \hat a_2^\dagger \hat a_1)  \nn \\
   && \quad + \frac{\bar g}{2}(\hat n_1(\hat n_1-1) + \hat n_2(\hat n_2-1)),
  \label{eqn-mp-hamiltonian}
\end {eqnarray}
where $\hat a_j$ and $\hat a_j^\dagger$ are the bosonic annihilation and
creation operators in the $j$th well and $\hat n_j = \hat a_j^\dagger\hat a_j $
is the occupation number operator.
The eigenvalues of the Hamiltonian (\ref{eqn-mp-hamiltonian}) are shown in
Fig.~\ref{fig-levels_mp_mf} on the right-hand side in dependence of
$\epsilon$ for $g=-1$, $v = 0.2$ and $N = 20$ particles.
One recognizes the similarity to the mean-field results shown on the
left-hand side. A series of avoided crossings with very small level
distances is observed where the mean-field energy levels form the
swallow's tail structure.

For $t \rightarrow -\infty$, one has $\epsilon = \alpha t \rightarrow - \infty$
and the first term dominates the Hamiltonian. The ground state is
$|\psi_0\rangle = (N!)^{-1/2} (\hat a_1^\dagger)^N |0\rangle$,
where $N$ is the fixed number of particles.
In the spirit of the Landau-Zener problem we take this as the initial
state for $t \rightarrow -\infty$ and consider the question, how many
particles remain in the first well  for $t \rightarrow +\infty$,
i.e. the effective Landau-Zener transition probability for the
{\it population}, which is given by
\be
  P_{\rm LZ}^{\rm mp} = \frac{\langle \hat n_1(t \rightarrow + \infty)\rangle}{\langle
  \hat n_1(t \rightarrow - \infty)\rangle} \, .
  \label{eqn-plz-mp-def}
\ee
The superscripts mp and mf are introduced to distinguish between the
many-particle and the mean-field system.
It will be shown that this many-particle Landau-Zener probability agrees
well with the mean-field Landau-Zener probability (\ref{eqn-plz-mf-def}).
Furthermore this ''back-to-the-roots''-procedure reduces the problem
to a {\rm linear} multi-level Landau-Zener scenario, which can be solved
approximately in an independent crossing approximation.
In this way we derive a Landau-Zener formula for an interacting BEC,
which agrees well with numerical results especially in the strongly
interacting regime $|g| > g_c = 2v$.

\section{The many-particle Landau-Zener problem and the ICA}
\label{sec-mp-ica}

We now consider the many-particle Landau-Zener scenario (\ref{eqn-mp-hamiltonian})
in detail, where the number $N$ of particles is fixed.
We expand the Hamiltonian $H$ in the number-state basis
$| k \rangle = [k!(N-k)!]^{-1/2} (\hat a_1^\dagger)^{k} (\hat a_2^\dagger)^{N-k} | 0 \rangle$.
Then the Hamiltonian is given by the matrix
$\langle \ell | H | k \rangle = H_ {\ell,k}$ for $\ell,k = 0, \ldots,N$
with the elements
\be
  H_{\ell,k} =  h_\ell(t) \, \delta_{\ell,k} + v_\ell \, (\delta_{\ell,k-1} + \delta_{\ell-1,k})
  \label{eqn-ham-matrix}
\ee
and
\[
   h_\ell(t) = \epsilon(t) (2\ell-N) + \frac{\bar g}{2} ( 2\ell^2 -2\ell N  + N^2 -N )
\]
and the couplings $v_\ell = v \sqrt{(\ell+1) (N-\ell)}$
on the sub- and superdiagonal.
In the Landau-Zener scenario, all diabatic (i.e. uncoupled) levels
$h_\ell(t)$ vary linearly in time as $\epsilon(t) = \alpha t$,
however with a different offset and slope $\alpha (2 \ell - N)$.

\begin{figure}[t]
\centering
\includegraphics[width=7cm,  angle=0]{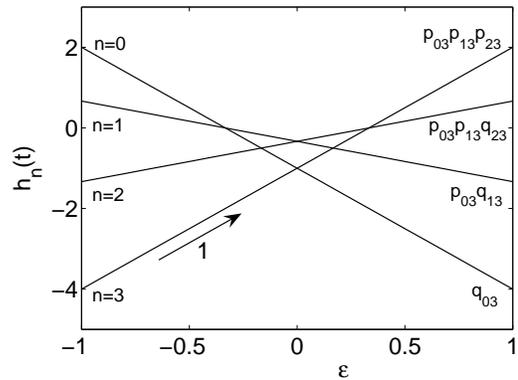}
\caption{\label{fig-levels_N=3}
The S-matrix elements $|S_{\ell,N}|^2$ in the independent crossing
approximation (ICA) for $N=3$ particles.}
\end{figure}

As stated above, we assume that initially all particles are in the
first well, $|\psi (t \rightarrow - \infty) \rangle = | N \rangle$.
Consequently one has $\langle \hat n_1(t \rightarrow - \infty)\rangle = N$
and in order to derive the Landau-Zener probability (\ref{eqn-plz-mp-def})
we are left with the problem to calculate
$\langle \hat n_1(t \rightarrow + \infty)\rangle$.
Thus we are not interested in the details of the time evolution.
We just need a few elements of the $S$-matrix, which is defined by
\be
   \label{eqn-S-matr-def}
   \langle k | \psi(t=+\infty) \rangle =
   \sum_\ell S_{k \ell} \langle \ell | \psi(t=-\infty) \rangle.
\ee
With this definition and $\langle \ell | \hat n_1 | k \rangle = k \delta_{\ell,k}$,
the Landau-Zener transition probability (\ref{eqn-plz-mp-def}) is reduced to
\be
  P_{\rm LZ}^{\rm mp} =  \frac{1}{N} \sum_{k=0}^{N} k |S_{k,N}|^2 \, ,
\ee
so that only the squared modulus of the $S$-matrix elements $|S_{k,N}|^2$
are of importance.

The S-matrix elements $|S_{N,k}|^2$ are now evaluated in a modified
independent crossings approximation (ICA, see appendix for details).
One assumes that the system undergoes
a series of single, independent transitions between just two levels.
The probabilities of a diabatic resp. adiabatic transition at a single
anti-crossing are given $p_{k,N} = \exp(-\pi w_{k,\ell}^2 / |b_{k,l}|)$
resp. $q_{k,\ell} = 1- p_{k,\ell}$ according to the Landau-Zener formula
(\ref{eqn-plz-lin}).
Here, $w_{k,l}$ denotes the level spacing at the anti-crossing and
$b_{k,\ell}$ is the difference of the slopes of the two diabatic levels.
The relevant S-matrix elements are given by
\be
  |S_{k,N}|^2 =  (1-p_{k,N}) \prod_{\ell = 0}^{k-1} p_{\ell,N} \, ,
\ee
with the definition $q_{NN} = 1 \Leftrightarrow p_{NN} = 0$.
The calculation of the S-matrix elements by the ICA is illustrated
in Fig.~\ref{fig-levels_N=3} for the case $N=3$.

\begin{figure}[t]
\centering
\includegraphics[width=7cm,  angle=0]{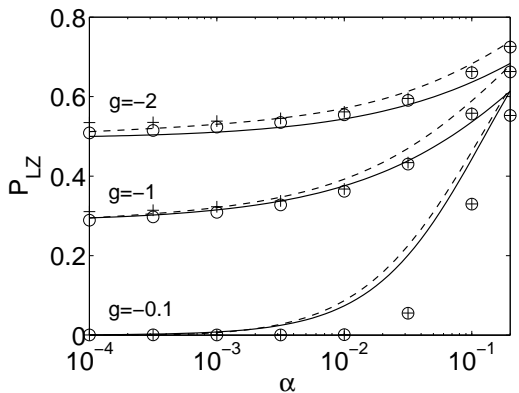}
\caption{\label{fig-plz-ica-avar} \label{fig-plz-avar2}
Landau-Zener tunneling probability in dependence of the parameter
velocity $\alpha$ for $v=0.2$, $N=100$ particles and different values of
the interaction constant $g$.
Numerical data (mean-field $+$ and many-particle theory $\circ$) is
compared with the ICA  (\ref{eqn-plz-ica}),(dashed line)
and the resulting ICA-Landau-Zener formulae (\ref{eqn-res-crit})
resp. (\ref{eqn-res-subcrit}),(solid line).}
\end{figure}

The ICA-Landau-Zener transition probability is then given by
\begin{eqnarray}
  P_{\rm LZ}^{\rm ICA} &=&  \frac{1}{N} \sum_{k=0}^{N} k (1-p_{k,N})
  \prod_{\ell = 0}^{k-1} p_{\ell,N} \nn \\
  &=&  \frac{1}{N} \sum_{k=0}^{N-1} \prod_{\ell=0}^{k} p_{\ell,N} \, .
  \label{eqn-plz-ica}
\end{eqnarray}
Note that the $p_{\ell,N}$ depend on the distance between the levels $\ell$
and $N$ at the anti-crossing. Thus they have to be evaluated at different
times $t_{\ell,N}$. However, the crossing time is easily calculated by evaluating
$h_N(t_{\ell,N}) =  h_\ell(t_{\ell,N})$, where $h_\ell(t)$ are diabatic
levels as defined above. This yields
\be
  t_{\ell,N} = -\frac{\bar g \ell}{2 \alpha} \, .
\ee
At all crossing times $t_{\ell,N}$, the level spacings $w_{\ell,N}$
are calculated by diagonalizing the Hamiltonian matrix (\ref{eqn-ham-matrix}).
As $H$ is tridiagonal, this can be done very efficiently.
Furthermore, the difference of the slopes is simply given by
$b_{\ell,N} =  2 \alpha (N-\ell)$.

To test this approach we compare the ICA-Landau-Zener formula (\ref{eqn-plz-ica})
with the Landau-Zener probability (\ref{eqn-plz-mp-def}) calculated
by numerically integrating the many-particle Schr\"odinger equation
as well as the mean-field transition probability (\ref{eqn-plz-mf-def}).
The results are shown in Fig.~\ref{fig-plz-ica-avar} in dependence of
$\alpha$ for $v = 0.2$, $N=100$ and three different values of $g$.
One observes a good agreement between the Landau-Zener
formula (\ref{eqn-plz-ica}),(dashed line) and the numerical 
results for large $g$.
For small values of $g$ the ICA (\ref{eqn-plz-ica}) overestimates
the transition probability. These issues will be further discussed
in section \ref{sec-lz-formula}.

\section{Limiting cases}

The linear limit $g \rightarrow 0$ is analytically solvable in both cases.
The many-particle system (\ref{eqn-ham-matrix}) reduces to the so-called
bow-tie model, whose S-matrix was calculated in \cite{Demk01}.
The mean-field dynamics reduces to the ordinary two-state model
of Landau, Zener, Majorana and St\"uckelberg \cite{Land32,Zene32,Majo32,Stue32}.
Not only the transition probability but also the whole dynamics
is known exactly in terms of Weber functions \cite{Zene32}.

\begin{figure}[t]
\centering
\includegraphics[width=7cm,  angle=0]{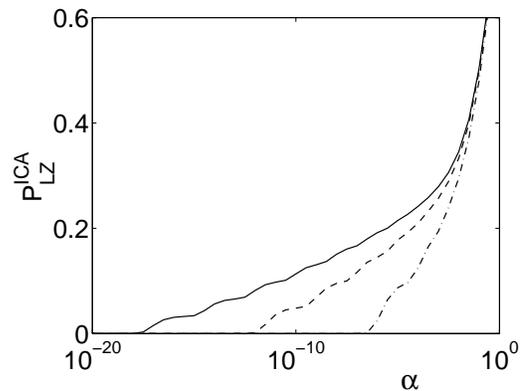}
\caption{\label{fig-plz-ica-adlim}
Landau-Zener transition probability $P_{\rm LZ}^{\rm ICA}(\alpha)$ in the
adiabatic limit $\alpha \rightarrow 0$ for $v = 0.2$, $g=-1$ and
different numbers of particles: $N=10$ ($-\cdot-$),
$N=20$ ($--$) and $N=30$ (---).}
\end{figure}

In the zero-coupling limit $v \rightarrow 0$ the Hamiltonians become diagonal
and the evolution is fully diabatic. The Landau-Zener transition probability
tends to one.

Most interesting is the adiabatic limit $\alpha \rightarrow 0$.
As no subdiagonal element of the Hamiltonian matrix (\ref{eqn-ham-matrix})
vanishes, all eigenvalues must be distinct (see, e.g. \cite{Wilk65}).
They may become pathologically close, but they cannot be degenerate.
This is in fact the case: The splitting of the lowest levels
at the anti-crossings becomes really small for increasing $|\bar g|$.
Thus all $w_{l,N}$ are non-zero and in the extreme adiabatic limit
$\alpha \rightarrow 0$ the Landau-Zener probabilities $p_{\ell,N}$
must vanish.
This seems to contradict the mean-field results
(cf. Fig.~\ref{fig-plz-ica-avar}), which predicts a non-zero Zener
tunneling probability even in the adiabatic limit if $|g| > g_c$.

However, the parameter regime, where the ICA predicts a vanishing
Zener tunneling probability in contrast to the mean-field results,
decreases rapidly with an increasing number of particles $N$.
Figure  \ref{fig-plz-ica-adlim} shows the Landau-Zener probability
$P_{\rm LZ}^{\rm ICA}(\alpha)$ for very small $\alpha$, calculated within
the ICA for different $N$, with $g = \bar g N = -1$ fixed.
The truly adiabatic region, where $P_{\rm LZ}^{\rm ICA}(\alpha) \approx 0$ is
negligibly small already for these quite modest numbers of $N$.

The mean-field theory is valid for a BEC consisting of a {\it macroscopic}
number of atoms. In order to compare to the mean-field results we
thus have to consider the limit of a large number of particles, 
$N \rightarrow \infty$ with $g = \bar g N$ fixed. In this macroscopic limit, 
the contradiction vanishes.
Furthermore, this limit will prove itself as extremely convenient for the
evaluation of Eq.~(\ref{eqn-plz-ica}), since all sums can be replaced
by integrals which can be solved explicitly (cf. section \ref{sec-lz-formula}).

\section{The many-particle spectrum}
\label{sec-spectrum}

The only missing step towards an explicit Landau-Zener formula is the evaluation
of the squared levels spacings $w^2_{k,N}(t_{k,N})$. Thus one has to understand the spectrum
of the Hamiltonian (\ref{eqn-mp-hamiltonian}). We start with a discussion of the
spectrum for $\epsilon = 0$, which provides an insight into the qualitative features
which will guide us in the following.
To keep the calculations simple, we introduce the operators
\begin{eqnarray}
  J_x &=& \frac{1}{2} \left(a_2^\dagger a_2 - a_1^\dagger a_1 \right) \nn \\
  J_y &=& \frac{\ri}{2} \left(a_2^\dagger a_1 - a_1^\dagger a_2 \right) \nn \\
  J_z &=& \frac{1}{2} \left(a_1^\dagger a_2 + a_2^\dagger a_1 \right),
\end{eqnarray}
which form an angular momentum algebra with quantum number $j = N/2$ \cite{Milb97,Vard01b,Angl01}.
The Hamiltonian (\ref{eqn-mp-hamiltonian}) then can be rewritten as
\be
  H =  2v J_z + \frac{g}{N} J_x^2
\ee
up to a constant term.

\begin{figure}[t]
\centering
\includegraphics[width=7cm,  angle=0]{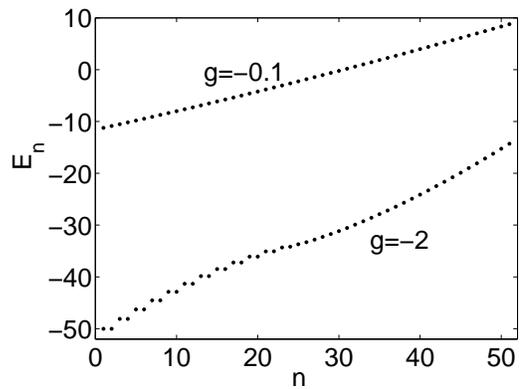}
\caption{\label{fig-spec-e0}
Spectrum of the many-particle Hamiltonian (\ref{eqn-mp-hamiltonian})
for $\epsilon = 0$, $v=0.2$, $N=50$ particles and $g=-0.1$ and $g=-2$,
respectively.}
\end{figure}

In the subcritical case $|g| < 2|v|$, the interaction terms can be treated
as a small perturbation. The unperturbed eigenstates are the
$J_z$-eigenstates $|j,m_z\rangle$ with $m_z = -j,-j+1,\ldots,j$.
In second order this yields the levels
\be
  E_{m_z} = 2vm_z \left[ 1 - \frac{g}{2v} \frac{m_z}{2N} - \left(\frac{g}{2v}\right)^2
   \frac{m_z^2}{4N^2} + \mathcal{O}(g^3) \right]
   \label{eqn-spec-pert1}
\ee
up to a constant. 
This spectrum is illustrated in Fig.~\ref{fig-spec-e0} for
$g=-0.1$, $v=0.2$ and $N=50$ particles. The eigenenergies
are nearly equidistant, with a slight increase of the level 
spacing for higher energies. 

For $|g| > 2|v|$ and low energies, the interaction term  $g  J_x^2/N^2$
dominates the Hamiltonian. The eigenstates with quantum numbers  $|j,\pm m_x\rangle$
are doubly degenerate with eigenenergy $E_{m_x} = g m_x^2/N^2$.
The perturbation $2 v J_z$ removes this degeneracy only in the $2|m_x|$-th
order. Thus the low energy eigenstates (corresponding to the high
$|m_x|$ states) appear in nearly degenerate pairs.
However this approach fails if the energy scale of the perturbation
$2 v J_z$ becomes comparable to the unperturbed eigenenergy.
Estimating the energy scale of the perturbation as $E_{\rm max}/2 = 2 v j/2 $,
perturbation theory fails for $|g| m_x^2/N^2 \apprle vN/2$.
Instead, Bogoliubov theory provides the appropriate description
for the high energy part of the spectrum.
We are dealing with an attractive interaction $g<0$,
so that the highest state in the mean-field approximation is the state
with equal population in the two modes. So the standard Bogoliubov
approach is valid for the highest state instead of the ground state.
One finds that the high energy part of the
spectrum is given by $E_n = E_N - \omega (N-n)$ with the Bogoliubov
frequency \cite{Pita03}
\be
  \omega = [(2v)^2 -2vg]^{1/2}. 
  \label{eqn-Bog-freq}
\ee
To clarify this issue, the spectrum is plotted in Fig.~\ref{fig-spec-e0}
for $g=-2$, $v=0.2$ and $N=50$ particles.
One clearly sees the nearly degenerate pairs of eigenvalues
for low energies and the approximately equal spacing of the
high-energy eigenvalues. The distance of the two highest
levels is given by the Bogoliubov frequency (\ref{eqn-Bog-freq}).

\begin{figure}[t]
\centering
\includegraphics[width=7cm,  angle=0]{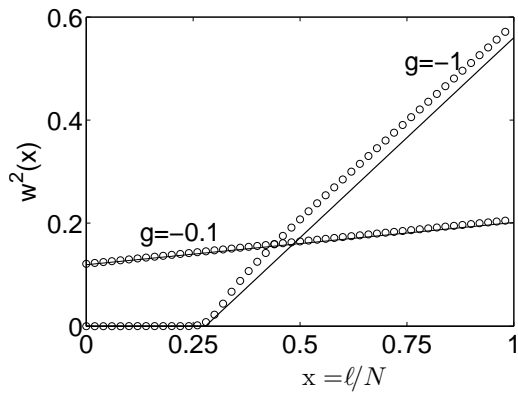}
\caption{\label{fig-w2-v=0.2}
Squared level splitting $w^2(x)$ in dependence of the scaled index
$x = \ell/N$ for $v =0.2$ and $g=-0.1$ resp. $g=-1$.
Numerical results ($\circ$) are compared to the approximate
formulae (\ref{eqn-w2-supercrit}) resp. (\ref{eqn-w2-subcrit}),
(solid lines).}
\end{figure}

Now we come back to the squared level splittings $w_{k,N}^2(t_{k,N})$,
beginning with the supercritical regime $|g| > 2v$.
Figure  \ref{fig-w2-v=0.2} shows an example of the squared level
splitting for $v=0.2$, $N = 100$ particles and $g=-0.1$ resp.
$g=-1$.
Later, we consider the macroscopic limit
$N \rightarrow \infty, \; \bar g \rightarrow 0$ with $g = \bar g N$
fixed. For this issue we plot the squared level splittings versus
the rescaled index $x := \ell/N \in [0,1]$.
With increasing $N$, the curve plotted in Fig.~\ref{fig-w2-v=0.2}
remains {\it the same}, only the actual points move closer together.
Thus one obtains a continuous function $w^2(x)$ in the limit
$N \rightarrow \infty$.

As argued above for $\epsilon = 0$, the lower levels appear in
approximately degenerate pairs. By the same arguments one concludes
that this is also true for for the first level crossings. Thus,
$w^2_{\ell,N}$ is effectively zero for $\ell < \ell_c$ resp.
$x < x_c$. The critical index $x_c$ can be estimated as described
above for $\epsilon = 0$. It is found that this estimate gives
the correct results up to a numerical factor $a$ of order 1. Thus we
conclude that
\be
  x_c \approx 1 - a \sqrt{2v/|g|}.
  \label{eqn-xc}
\ee
A very good agreement of this formula to the numerical results
was found for $a = 1.14$.

For $x > x_c$ the squared splittings increase approximately linear.
In the high energy limit corresponding to $x \rightarrow 1$, the level
splitting is given by the Bogoliubov frequency introduced above.
In conclusion, the squared level spacing can be approximated by
\be
  w^2(x) \approx \omega^2 \frac{x-x_c}{1-x_c} H(x-x_c),
  \label{eqn-w2-supercrit}
\ee
where $H(x-x_c)$ denotes Heaviside's step function.

In the subcritical regime $|g| < 2v$, one can use the results from
perturbation theory described above (cf. Eq.~(\ref{eqn-spec-pert1})).
At time $t_{\ell,N} = -\bar g \ell/2\alpha$ one must evaluate the level
splitting $E_{\ell-j+1} - E_{\ell-j}$ (note that the levels are
labeled by $m_z = -j,-j+1, \ldots, j$ with $j = N/2$). Again we consider
the limit $N \rightarrow \infty$ with $g = \bar g N$ fixed.
After a little algebra one finds that the relevant level
splitting is in linear order given by
\begin{eqnarray}
  w(x) &=& \frac{16v^2+4g - 3g^2/4}{8v} +
  \left( \frac{3 g^2}{8v} - g \right) x  \nonumber \\
    &=:& w_0 + w_1 x
    \label{eqn-w2-subcrit}
\end{eqnarray}
in terms of the scaled index $x = \ell/N$.

The approximate results for the squared level splitting $w^2(x)$
for $|g| > g_c$, Eq.~(\ref{eqn-w2-supercrit}), and for $|g| < g_c$,
Eq.~(\ref{eqn-w2-subcrit}), are compared with the numerical results
for $N=100$ particles in Fig.~\ref{fig-w2-v=0.2}. One observes
a good agreement.

\section{Towards an explicit Landau-Zener formula}
\label{sec-lz-formula}

Using the formulae for the squared level splitting derived in the
previous section, the ICA-Landau-Zener transition probability
(\ref{eqn-plz-ica}) can now be evaluated explicitly.
In the spirit of the of the macroscopic limit $N \rightarrow \infty$,
the sums are replaced by integrals according to
\be
  \frac{1}{N} \sum_{\ell = 0}^{k} \rightarrow \int_0^{k/N} dx
\ee
The difference of the slopes $b_{\ell,N} =  2 \alpha (N-\ell)$,
which enters the formula is also rewritten in terms of the rescaled
index $x = \ell/N$:
\be
   b_{\ell,N} \rightarrow 2 \alpha N (1-x) =: N \bar b(x).
   \label{eqn-barb}
\ee
Thus one finds
\be
   P_{LZ} \approx \int_0^1 \exp\left[ -\pi \int_0^y \frac{w^2(x)}{\bar b(x)} dx \right] dy.
   \label{eqn-plz-intapp}
\ee

In the supercritical regime $|g| > g_c$ we start by evaluating
the integral over $x$ in Eq.~(\ref{eqn-plz-intapp}).
Substituting $w^2(x)$ and $\bar b(x)$ from Eq.~(\ref{eqn-w2-supercrit})
and (\ref{eqn-barb}) and carrying out the integral yields
\be
  \int_0^y \frac{w^2(x)}{\bar b (x)} dx = \frac{-\omega^2}{2 \alpha}
  \left[ \frac{y-x_c}{1-x_c} + \ln\left(\frac{1-y}{1-x_c}\right) \right]
\ee
for $y > x_c$ and zero otherwise.
The Landau-Zener transition probability (\ref{eqn-plz-intapp})
is then given by
\begin{eqnarray}
  P_{\rm LZ} &\approx& x_c + \int_{x_c}^1
  \left( \frac{1-y}{1-x_c} \right)^{ \frac{\pi \omega^2}{2 \alpha}}
  \exp \left[ \frac{\pi \omega^2}{2 \alpha} \frac{y-x_c}{1-x_c} \right] dy \nonumber \\
  &=& x_c + \frac{(1-x_c) \, \re^u}{u^{u+1}} \, \gamma(u+1,u)
  \label{eqn-res-crit}
\end{eqnarray}
with the abbreviation $u = \pi \omega^2 / 2\alpha$
and $x_c$ defined in Eq.~(\ref{eqn-xc}). Here, $\gamma$
denotes the incomplete gamma-function \cite{Abra72}.

In the subcritical regime $|g| < g_c$, one finds by substituting
Eq.~(\ref{eqn-plz-intapp}) into (\ref{eqn-plz-intapp}), that
the Landau-Zener transition probability is given by
\begin{eqnarray}
  && P_{LZ} \approx \int_0^1 \left( 1-y \right)^{(\pi w_0 (w_0+2 w_1))}
  \exp\left[2\pi w_0 w_1 y \right] dy \nonumber \\
  && \; = \frac{\re^{c_1}}{c_1^{c_0+1}}
  \, \gamma \left(c_0+1, c_1\right)
    \label{eqn-res-subcrit}
\end{eqnarray}
with the abbreviations $c_0 = \pi (w_0^2 +2 w_0 w_1) /2 \alpha$ and
$c_1 = \pi w_0 w_1/\alpha$. 
To keep the calculations feasible, we kept only terms linear in $x$ resp. $y$
in the exponent consistent with Eq.~(\ref{eqn-res-crit}).

\begin{figure}[t]
\centering
\includegraphics[width=7cm,  angle=0]{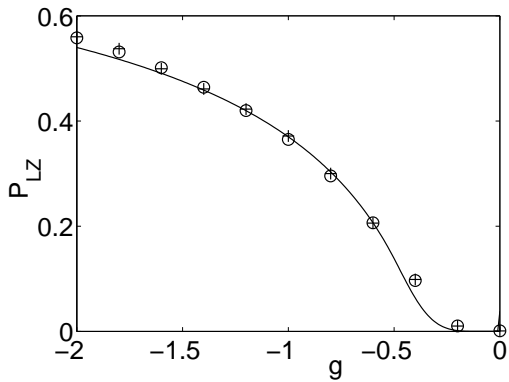}
\caption{\label{fig-plz-gvar2}
Landau-Zener tunneling probability in dependence of the interaction
constant $g$ for a parameter velocity $\alpha = 0.01$.
Numerical data (mean-field $+$ and many-particle theory $\circ$) is compared
to the ICA-Landau-Zener formula (\ref{eqn-res-crit}),(solid line) for $v=0.2$.}
\end{figure}

To test the validity of our approach we compare the
ICA-Landau-Zener formulae (\ref{eqn-res-crit}) and (\ref{eqn-res-subcrit})
to numerical results obtained by integrating the Schr\"odinger
equation for mean-field Hamiltonian (\ref{eqn-ham-nonlin})
as well as the many-particle Hamiltonian (\ref{eqn-mp-hamiltonian}).
The Landau-Zener tunneling probability in dependence of the interaction
constant $g$ is plotted in Fig.~\ref{fig-plz-gvar2} for
$\alpha = 0.01$ in dependence of the velocity parameter $\alpha$
for different values of $g$ in Fig.~\ref{fig-plz-avar2}.

One observes a good agreement of the ICA-Landau-Zener formula with
the numerical results in the critical regime $|g| > g_c$.
Especially the increase of the tunneling probability with
increasing $|g|$ for small $\alpha$ is well described by our model.
This problem could not be solved with previous approaches
\cite{Zoba00,Liu02}.
The approximation gets worse for larger values of $\alpha$ because
the assumption that the Zener transitions are well separated
becomes doubtful for such a large parameter velocity. The ICA thus
underestimates the tunneling probability.

In the subcritical case $|g| < g_c$, the proposed ICA-Landau-Zener formula
does not work as well. In fact the tunneling probability is overestimated
for small $\alpha$ because the ICA itself is not a very good approximation
in this case. The adiabatic levels do not show well separated avoided
crossings, instead the levels splittings are nearly constant over a long
interval of the parameter $\epsilon$.
For larger values of $\alpha$ one faces the same problems as in the
supercritical case and the tunneling probability is underestimated.
Another ansatz, using e.g. perturbation theory with respect to the
solution of the noninteracting problem \cite{Angl03} should be better
suited to this problem. Note however, that the deviations are mainly
due to the ICA itself and to the approximation of $w^2(x)$ made in this
section.

\section{Conclusion and Outlook}

In conclusion, we have derived a Landau-Zener formula for an interacting
Bose-Einstein condensate from first principles. To this end we considered
the original two-mode many-particle Landau-Zener scenario.
It was shown that the resulting Landau-Zener formula agrees well with the
numerical results calculated for the many-particle problem as well as within
the mean-field approximation.

In the future, it would be of interest to relate our calculations to the
respective problem in the Heisenberg pictures. Here, complex eigenfrequencies
may occur for the dynamics of the creation/annihilation operators, leading to
spontaneous production of quasi-particles and hence a dynamical instability.
For the non-interacting case, this problem has been solved analytically
\cite{Angl03}.

Another issue is the discussion of nonlinear Landau-Zener problems for more
than two levels. First results for three level system were reported only
recently \cite{05level3}.

\section*{Appendix: The independent crossings approximation}

\begin{figure}[thb]
\centering
\includegraphics[width=7cm,  angle=0]{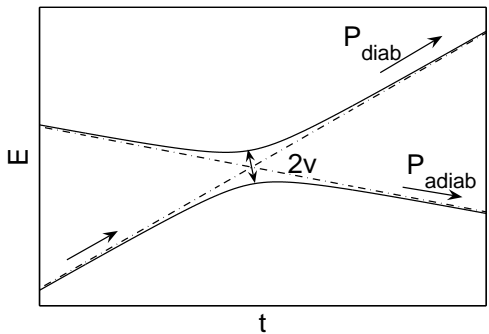}
\caption{\label{fig-P_LZlin}
Diabatic (dash-dotted line) and adiabatic (solid line) energy levels
of the two-level Landau-Zener model (\ref{eqn-lin2lev}).}
\end{figure}

Let us first briefly recall the dynamics of a two-level Landau-Zener system
described by the Hamiltonian
\be \label{eqn-lin2lev}
  H_0(t)=\left(\begin{matrix}
  \beta_1 t+b_1 & v\\
  v & \beta_2 t+b_2
  \end{matrix}\right).
\ee
The diabatic and adiabatic energy curves are plotted in Fig.~\ref{fig-P_LZlin}.
The S-Matrix in the sense of Eq.~\ref{eqn-S-matr-def} is given by
\be
  S=\left(\begin{matrix}
  p & q\\
  q & p
  \end{matrix}\right)
\ee
with $p=\exp{\left(-\pi v^2/\vert \beta_1-\beta_2\vert \right)}$ and $q=\sqrt{1-p^2}$.
The probability of a diabatic passage is therefore given by the Landau-Zener-formula
$P_{\rm LZ}=p^2= \exp{\left(-\frac{2\pi v^2}{\vert \beta_1-\beta_2 \vert}\right)}$,
and the adiabatic transition probability by $1-P_{\rm LZ}=q^2$. They depend only on
the relative slope of the diabatic levels $\vert \beta_1-\beta_2\vert$ and the
coupling $v$, which is equivalent to half of the gap between the adiabatic
energy-levels at the avoided crossing.

The simplicity of the solution of the two-level system and the observation that
the transitions between two adiabatic levels in a multilevel Landau-Zener system
takes place only in a very narrow region around the crossing of the two corresponding
diabatic levels leads to a simple approximation. If all crossings are well
separated they can be considered as independent of each other and each of them
is described by the two-level-Landau-Zener model where the couplings between the
relevant diabatic levels are given by the nondiagonal terms of the Hamiltonian.
This approach is called the "independent crossing approximation" (ICA) in the
literature. It is of great importance for the study of multilevel Landau-Zener
dynamics because of a surprising feature: The ICA turns out to give the {\it exact}
results for all known exactly solvable multi-level Landau-Zener scenarios \cite{Demk68,Demk01}. Furthermore it has been shown that the ICA always gives the correct results for the
diagonal $S$-matrix elements with minimal and maximal slope \cite{Sini04,Shyt04}.

\begin{figure}[thb]
\centering
\includegraphics[width=7cm,  angle=0]{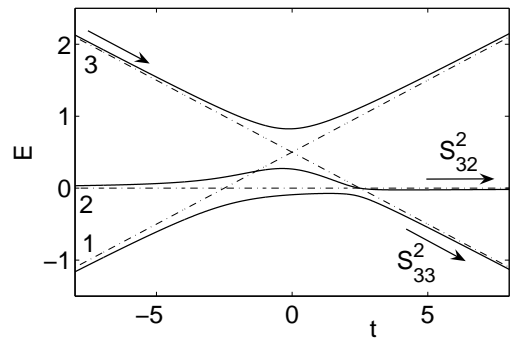}
\caption{\label{fig-3niv}
Diabatic (dash-dotted line) and adiabatic (solid line) energy levels
of the three-level Landau-Zener model (\ref{eqn-ica-3niv}) for
$\alpha = 0.2$, $a=0.5$, $v=0.2$ and $w=0.3$.}
\end{figure}

Of course there are also examples where the ICA fails, as for example for the
simple three level Hamiltonian
\be
H(t) = \left(\begin{matrix}
  \alpha t+a & v & w\\
  v & 0& 0\\
  w & 0 & -\alpha t+ a
  \end{matrix}\right).
  \label{eqn-ica-3niv}
\ee
The adiabatic and diabatic levels are plotted in figure \ref{fig-3niv}.
The diabatic transition probability for the third level $S_{33}$ is exactly
given by the ICA. But if we look at the S-Matrix element $S_{32}$ we find that
the ICA predicts it to be zero, because the coupling matrix element
vanishes, $\langle 2\vert H\vert 3\rangle=0$, independent of $\alpha$
and $a$, which isn't true.
The second and the third diabatic levels do not couple directly, but for not too
large values of $a$ the indirect coupling via the second diabatic level can't be
neglected. This coupling manifests itself as an avoided crossing between the two
adiabatic levels, which turns into a real crossing only in the limits $a\to\infty$
and $a\to 0$. For finite values of $a$ the transition probability between the third
and the second diabatic levels is small but nonzero.

To get a better approximation one should recall the two level system, where the
coupling between two diabatic levels is equivalent to half of the level-splitting
of the corresponding adiabatic levels. Therefore one can use a modified ICA where
the couplings are not given by the nondiagonal elements of the Hamiltonian but half
of the level splitting between the relevant adiabatic levels. This approximation
doesn't inherit the benefit of providing the exact results in the special cases
where the original ICA did, but provides a good approximation even in the cases
where the ICA fails.
Therefore it is better suited for our purposes. The performance of the approximation
is limited by the fact that the single avoided crossings must be well separated so
that the transition regimes do not overlap. In the present case this is improved
with increasing nonlinearity. Note that to simplify matters this modified ICA is denoted
as ICA throughout the paper.

\begin{acknowledgments}
Support from the Studienstiftung des deutschen Volkes and the Deutsche
Forschungsgemeinschaft via the Graduiertenkolleg ''Nichtlineare Optik
und Ultrakurzzeitphysik'' is gratefully acknowledged.
\end{acknowledgments}


\end{document}